\documentclass[12pt]{article}
\usepackage{epic,eepic}
\usepackage{exscale}
\begin{document}
\hfill LMU 16/04
\begin{center}
{\large \bf The Spin Structure of the Constituent Quarks and of
the Nucleon}
\end{center}
\bigskip
\begin{center}
{\bf Harald Fritzsch}
\end{center}
\begin{center}
Arnold Sommerfeld Center, University of Munich\\
Theresienstra{\ss}e 37, D-8033 M\"unchen
\end{center}
\noindent
\begin{abstract}
\noindent
We define a constituent quark within QCD. It is shown that the spin of
such a quark and hence also the spin of the nucleon reduced due to
$\bar q q$--pairs, in agreement with experiment. A solution to the spin
problem is given.
\end{abstract}
Experiments have revealed that the internal structure of the nucleon is more
complicated than originally assumed$^{1)2)}$. In particular, the portion of
the nucleon spin carried by the spins of the u- and d-quarks is not, as
naively expected, about 75\%, but much smaller. In this paper we study the
situation by considering the internal spin structure of the constituent quarks.

Often it is assumed that the nuleon consists of three constituent quarks,
each of them having their own internal structure. However, in QCD the
notion of a constituent quark has remained vague. Using a specific
``gedankenexperiment'' we first show how a constituent quark can be
defined with all its dynamical properties. Then we apply our results to
the nucleon. In order to simplify the task, we shall first neglect the
s-quarks. Later we comment on what changes once s-quark are introduced.

We relate the spin structure to the QCD anomaly$^{3)}$. A very specific
picture emerges, giving an elegant solution of the spin problem.

For a proton being a system of three constituent quarks, i. e. (uud), it
is difficult to disentangle the contributions of the three constituent
quarks. Therefore we consider a heavy baryon of spin 3/2, e. g. one with a
quark structure (bbu). The ground state of such a system is an
isospin doublet, which we denote as U and D:
\begin{eqnarray}
U\Uparrow & = & (b\Uparrow b\Uparrow u\Uparrow) \\
D\Uparrow & = & (b\Uparrow b\Uparrow d\Uparrow) \nonumber \, .
\end{eqnarray}
One expects that these states of spin 3/2 exist in reality, but will
probably never be observed and studied in detail. We proceed to study
the internal structure of these states, in particular the aspects related
to the light quarks. Note that a state like (bbu) consists of a single
light constituent quark.

The states $U$ and $D$ behave much like the proton--neutron system, if we
turn off the weak interaction of the $b$--quarks. For example, the
state $D$, being slightly heavier than $U$, would show a $\beta$-decay:
$D \rightarrow U + e^ - + \bar \nu_e$. This can be used to define the 
associated vector and axialvector coupling constants, e. g. the matrix
elements $< U | \bar u \gamma_{\mu} d | D >$ and $ < U | \bar u \gamma_{\mu}
\gamma_5 d | D >$.

The isospin doublet $ U - D $ would exhibit a strong interaction with
pions, e. g. there would be charge exchange reactions like
 $\pi^ - U \rightarrow \pi^ {\circ}D $. Relations analogous to the
Goldberger--Treimann relations or the Adler--Weissberger relations would
be valid.

Furthermore the $U$-- or $D$--state can be used in a ``gedankenexperiment''
as target for lepton scattering experiments. This way the distribution
functions of the light quarks $u, d$ can be studied. The heavy quark
$b$ would constitute essentially a fixed portion of the momentum of the
$U$-- or $D$--state. The associated distribution function would essentially
be a $\delta $--function in $x$--space. Thus the heavy quark contribution
to the total momentum can be disregarded. What is left over, is the
momentum distribution function of the constituent light quark, which we
would like to investigate.

The light quark distribution functions are then given in terms of a scaling
parameter $x$, defined to be the momentum of the quarks, divided by the
total momentum of the constituent quark. Thus the variable $x$ varies as
usual between zero and one.

The states $U$ and $D$ can be polarized. The simple $SU(6)$ wave function
is given by:
\begin{eqnarray}
U\Uparrow & = & \frac{1}{\sqrt{3}} \left[\left( b\Uparrow b\Uparrow u\Uparrow
\right) + \left( b\Uparrow u\Uparrow c
b\Uparrow \right) \right. \\
& & + \left. \left( u\Uparrow b\Uparrow b\Uparrow \right) \right] \nonumber 
\end{eqnarray}

In QCD the light quark distribution functions are given by the matrix
elements of the bilocal densities $\bar q(x) \gamma_{\mu} q(y)$ or
$\bar q(x) \gamma_{\mu} \gamma_5 q(y)$ at lightlike distances. Taking
these matrix elements, one arrives at the distribution functions
$u_+ (x), u_-(x), d_+(x)$ and $d_- (x)$ of the $U$--state. The indices +
or - denote the helicity + or - of the corresponding quark in a polarized
$U$--state with positive helicity.

Let us first denote the sum rules following from the exact flavor
conservation laws. The matrix element $< U | u^ + u | U >$ is, of course,
given by one, the matrix element
$< U | d^ + d | U >$ vanishes. Thus we have
\begin{eqnarray}
\int\limits^{1}_{0} \left( u_+ + u_- - \bar u_+ - \bar u_- \right) dx & = & 1\\
\int\limits^{1}_{0} \left( d_+ + d_- - \bar d_+ - \bar d_- \right) dx & = & 0
\nonumber 
\end{eqnarray}

These rules above are the analog of the Adler sum rule in the case
of nucleons. We proceed to discuss the analog of the Bjorken sum rule
denoting the axial coupling constant of the $U-D$--system by $g_a$:
\begin{equation}
\int\limits^{1}_{0} \left\{ \left[ \left( u_+ + \bar u_+ \right) -
\left( u_- + \bar u_- \right)\right] - \left[ \left( d_+ + \bar d_+ \right) -
\left( d_- + \bar d_- \right) \right] \right\} dx = g_a
\end{equation}
\\
This sum rule concerns the isotriplet, i. e. the matrix element
$< U | \bar u \gamma_{\mu} \gamma_5 u_- - \bar d \gamma_{\mu} \gamma_5 d |U >$.
We can also consider the matrix element of the isosinglet current
$\bar u \gamma_{\mu} \gamma_5 u + \bar d \gamma_{\mu} \gamma_5 d$. The
associated sum rule
\begin{equation}
\int^{1}_{0} \left[ \left( u_+ + \bar u_+ - u_- - \bar u_- \right) +
\left( d_+ + \bar d_+ - d_- - \bar d_- \right) \right] dx = \Sigma
\end{equation}
gives a number $\Sigma$, which can be viewed as the contribution of the
$u$-- and $d$--quarks to the $U$--spin.

In a naive model the constituent $u$--quark inside the $U$--particle would
be composed solely of a $u$--quark with positive helicity, i. e. all
density functions vanish except for $u_+$:
\begin{equation}
\int\limits^{1}_{0} u_+ dx = 1 \qquad d_+ = d_- = \bar d_+ = \bar d_-
= u_- = \bar u_+ = 0 \qquad g_a = \Sigma = 1
\end{equation}

This relation would correspond to the $SU(6)$-result in the nucleon:
$\left|\,  G_A / G_V \right| = 5/3$. In reality we have
$\left| G_A / G_V \right| = 1.26$, i. e. a reduction from 5/3 by nearly 25\%.
Taking the same reduction for $U$, $D$, as an example, we expect
for the $U / D$--system: $g_a \cong 0.75$ instead of $g_a = 1$.

The axialvector coupling constant $g_a$ would fulfill a Goldberger--Treiman
relation and would be related to the $\pi -U-D$-coupling constant. The
associated axialvector current will be conserved in the limit
$m_u = m_d = 0$.

However the singlet current $\bar u \gamma_{\mu} \gamma_5 u + \bar d 
\gamma_{\mu} \gamma_5 d$ is not conserved in this limit due to the gluon
anomaly:
\begin{equation}
\partial^{\mu} \left( \bar u \gamma_{\mu} \gamma_5 U + \bar d \gamma_{\mu}
\gamma_5 d \right) = \frac{g^ 2}{4 \pi^ 2} \cdot \frac{1}{8}
\varepsilon_{\alpha \beta \gamma \delta} G_a^{\alpha \beta}
G_a^{\gamma \delta} 
\end{equation}
($G_a^ {\mu \nu}$: Gluon field strength).

It is well--known that the gluonic anomaly leads to an abnormal mixing
pattern for the $0^ {-+}$--mesons, implying a strong violation of the
Zweig rule in the $0^ {-+}$--channel. We consider the anomaly as the
reason why in the case of the nucleon the axial singlet charge deviates
strongly from the naive quark model value$^{3)}$.\\
\\
As can be seen directly from the sum rules given above, we obtain
immediately the naive result $g_a = \Sigma$, if all $d$--densities vanish. If
we take as an example $g_a = 0.75$ and $\bar u = d = \bar d = 0$, we obtain:
\begin{eqnarray}
\int\limits^{1}_{0} \left( u_+ - u_- \right) dx & = & 0.75 \qquad
\int\limits^{1}_{0} \left( u_+ + u_- \right) dx = 1 \\
\int\limits^{1}_{0}  u_+ dx & = & 0.875  \qquad
\int\limits^{1}_{0}  u_- dx  = 0.125 \nonumber
\end{eqnarray}

In this case 75\% of the $U$--spin would be given by the spin of the
$u$--quark, the remaining 25\% are due to other effects like orbital
effects and gluons.

However in the presence of the QCD anomaly the picture changes since
$\Sigma \not= g_a$. We isolate the $d$--integral and obtain:
\begin{equation}
2 \int\limits^{1}_{0} \left( d_+ + \bar d_+ - d_- - \bar d_- \right) dx
= \Sigma - g_a
\end{equation}

The difference $\Sigma - g_a$ is given by the matrix element $< U | \bar d
\gamma_{\mu} \gamma_5 d | U >$ which in a naive picture vanishes. We
decompose this m. e. into an isosinglet and isotriplet term:
\begin{equation}
 < U | \bar d \gamma_{\mu} \gamma_5 d | U > = \frac{1}{2} < U |
\bar d \gamma_{\mu} \gamma_5 d - \bar u \gamma_{\mu} \gamma_5 d | U > \\
+ \frac{1}{2} < U | \bar d \gamma_{\mu} \gamma_5 d + \bar u
\gamma_{\mu} \gamma_5 u | U >
\end{equation}

The isospin triplet term is determined via a Goldberger--Treimann relation and
related to PCAC and the associated pion pole.

Suppose, PCAC would also be valid for the singlet current. In this case
there would be a Goldstone particle (the $\eta$--meson with quark composition
$\frac{1}{\sqrt{2}} \left( \bar u u + \bar d d\right)$, and the
$\pi^ {\circ}$ and $\eta$--contribution would cancel. The matrix element
vanishes, and we have $\Sigma = g_a$.

In reality this is not true. One finds, for example, for the nucleon
$\Sigma \approx 0.30$. We set as an illustration
$\Sigma = 0.30$ and obtain:
\begin{eqnarray}
\int\limits^{1}_{0} dx \left( d_+ + \bar d_+ - d_- -\bar d_-\right) & = &
\frac{1}{2} \left( \Sigma - g_a \right) \cong - 0.22 \\
\int\limits^{1}_{0} dx \left( u_+ + \bar u_+ - u_- -\bar u_-\right) & = &
\frac{1}{2} \left( \Sigma + g_a \right) \cong 0.53 \nonumber
\end{eqnarray}

We note that due to the QCD--anomaly $\bar q q$--pairs are generated inside
the $U$--particle. This is a nonperturbative effect, like to
QCD--anomaly itself. Furthermore the $\bar q q$--pairs are polarized,
cancelling partially the spin of the $u$--quark. Note that the sign
of $\left( \Sigma - g_a \right)$ is negative. The sum rule for the
$d$--densities above implies that the sum $\left( d_- + \bar d_- \right)$
is nonzero, but it does not imply that the sum $\left( d_+ + \bar d_+ \right)$
is nonzero. Thus $\left( d_+ + \bar d_+ \right)$ can be zero or very
small.

The $\left( \bar q q \right)$--pairs, generated by the QCD anomaly,
e. g. by the gluonic dynamics, are not related to the
$u$--quark directly, and one therefore expects in particular
$d_- = \bar d_-$. The simplest way to obey the sum rules is
$d_+ = \bar d_+ = 0, \, d_- = \bar d_- = \bar u_-$.

In this case a polarized constituent $u$--quark is dominated by the
$u_+$--function, accompanied by $\left( \bar d d \right)$-- and
$\left( \bar u u \right) $--pairs, which
partially cancel the spin of the $u$--quark.

An interesting case, probably close to reality is:
\begin{equation}
\bar u_+ = 0, \qquad d_+ = \bar d_+ = 0  
\end{equation}
\begin{eqnarray}
\int\limits^{1}_{0} \left( d_- + \bar d_- \right) dx & = &
- \frac{1}{2} \left( \Sigma - g_a \right) \cong 0.22 \nonumber \\
\nonumber \\
\int\limits_{0}^{1} \bar u_- dx & = & - \left( \Sigma - g_a
\right) \cong 0.11 \nonumber \\
\nonumber \\
\frac{1}{2} \left( \Sigma + g_a \right) & = & \int\limits^{1}_{0} dx
\left( u_+ - u_- - \bar u_- \right) \cong 0.53 \, . 
\end{eqnarray}
\\
Now we include the $s$--quark. At first we consider the case of
$SU(3)$--symmetry: $m_u = m_d = m_s$. The total spin sum is given by
\begin{equation}
\int\limits^{1}_{0} \left[ ( u_+ + \bar u_+ - u_- - \bar u_-)
 +  ( d_+ + \bar d_+ - d_- - \bar d_- ) +  ( s_+ + \bar s_+ - s_- - \bar s_-)
\right] dx = \Sigma
\end{equation}
\\
We obtain:
\begin{equation}
\int^{1}_{0} \left[ \left( d_+ + \bar d_+ - d_- - \bar d_- \right) 2
+ \left( s_+ + \bar s_+ - s_- - \bar s_- \right) \right] dx = 
\Sigma - g_a
\end{equation}
\\
Again we set $s_+ = \bar s_+ = d_+ = \bar d_+ = 0$. Furthermore we have
$d_- = \bar d_- = s_- = \bar s_-$, and obtain:
\begin{eqnarray}
\int \left( d_- + \bar d_- \right) dx & = & \int \left[ \left( s_- +
\bar s_- \right) \right] dx \cong 0.15 \nonumber \\
\nonumber \\
\int \bar u dx & \cong & 0.075
\end{eqnarray}
\\
Again the spin is partially cancelled by the $\bar q q$--pairs, this time
including $\bar s s$--pairs.\\
\\
Now we consider the realistic case with $SU(3)$--braking. It is well--known
that the physical wave function of the $\eta$--meson and the $\eta'$--meson
are approximately given by
\begin{equation}
\eta = \frac{1}{2} \left( \bar u u + \bar d d - \sqrt{2} \bar s s \right),
\quad
\eta' = \frac{1}{2} \left( \bar u u + \bar d d + \sqrt{2} \bar s s \right) \, .
\end{equation}
\\
Thus in reality we are between the two cases discussed above. As an
example, which is probably close to reality, we take
\begin{eqnarray}
\int \left( d_- + \bar d_- \right) dx & = & 0.18 \nonumber \\ \nonumber \\
\int \bar u_- dx & \cong & 0.09 \nonumber \\ \nonumber \\
\int \left( s_- + \bar s_- \right) dx & = & 0.11\, .
\end{eqnarray}
\\
Again we see that the QCD anomaly is the reason why the spin is partially
cancelled by the $\bar q q$-pairs, although the $\bar ss$--pairs are less
relevant than the $\bar d d$-- and $\bar u u$--pairs.\\
\\
We think that we have found in the
QCD anomaly the reason why the spin of the nucleon is a rather complicated
object. The spin is reduced by $\bar qq$--pairs, which partially cancel the
spin of the constituent quarks. The $\bar qq$--pairs are polarized. This
polarization can be observed in lepton--nucleon--scattering.\\
\\
The question arises who carries the remaining part of the spin. The
departure of $|G_A / G_V|$ from $5/3$ indicates that orbital effects are
there. They make up about 25\% of the spin of the nucleon. The remaining
part of about 45\% is related to the QCD--anomaly. Since the latter is a
$\bar qq$--effect, we conclude that about 45\% of the nucleon spin is
carried by gluons. We summarize: 30\% of the spin is carried by the
valence quarks and the $\bar q q$--pairs. 25\% by orbital effects, 45\% by
gluons.\\
\\
The polarized $\bar qq$--pairs should be searched for in the experiments.
Also the gluonic constribution can be observed in the experiments,
especially by studying the $\bar cc$--production in
lepton--nucleon--scattering, as done in the Compass experiment$^{4)}$.\\
\\
I am indebted to Prof. A. Faessler for discussions.\\
\\
\\
References
\begin{enumerate}
\item[1)] For review, see e. g.: J. R. Ellis and M. Karliner,
{\it hep--ph/9601280}
\item[2)] See e. g.: B. Adeva et. al., {\it Phys. Rev.}
{\bf D60}: 072004 (1999)
\item[3)] H Fritzsch {\it Phys. Lett} {\bf 229B} 122 (1989)\\
{\it Nucl. Phys.} {\bf B15} 261 (1990)\\
{\it  Phys. Lett.} {\bf B242} 451 (1990)\\
{\it Phys. Lett.} {\bf 256B} 75 (1990)\\
{\it Nucl. Phys.} {\bf 23B} 91 (1991)\\
\item[4)] G. Baum et. al., CERN.SPSLC -- 96 -- 14
\end{enumerate}
\end{document}